\documentclass[aps,prl,showpacs,twocolumn,superscriptaddress,preprintnumbers,floatfix]
{revtex4}

\usepackage{graphicx}
\usepackage{epsfig}
\usepackage{amsmath,amssymb}
\usepackage{psfrag}
\begin{document}

\newcommand{\be}{\begin{eqnarray}}
\newcommand{\ee}{\end{eqnarray}}
\newcommand\del{\partial}
\newcommand\nn{\nonumber}
\newcommand{\Tr}{{\rm Tr}}
\newcommand{\mat}{\left ( \begin{array}{cc}}
\newcommand{\emat}{\end{array} \right )}
\newcommand{\vect}{\left ( \begin{array}{c}}
\newcommand{\evect}{\end{array} \right )}
\newcommand{\tr}{\rm Tr}
\def\conj#1{{{#1}^{*}}}
\newcommand\hatmu{\hat{\mu}}

\title{Geometric Flows and Black Holes}

\author{Fu-Wen Shu}
\email{fwsu@shao.ac.cn} \affiliation{Shanghai Astronomical
Observatory, Chinese Academy of Sciences, Shanghai 200030, P. R.
China} \affiliation{Graduate School of Chinese Academy of Sciences,
Beijing 100039, P. R. China}
 \affiliation{Joint Institute for Galaxy and Cosmology of SHAO and USTC, Shanghai 200030, P. R.
China}
\author{You-Gen Shen$^{1,3,\,}$}
\email{ygshen@shao.ac.cn} \affiliation{National Astronomical
Observatories, Chinese Academy of Sciences, Beijing 100012, P. R.
China}

\date   {\today}

\begin  {abstract}
Motivated by the newest progress in geometric flows both in
mathematics and physics, we apply the geometric evolution equation
to study some black-hole problems. Our results show that, under
certain conditions, the geometric evolution equations satisfy the
Birkhoff theorem, and surprisingly, in the case of spherically
symmetric metric field, the Einstein equation, the Ricci flow, and
the hyperbolic geometric flow in vacuum spacetime have the same
black-hole solutions, especially in the case of $\Lambda=0$, they
all have the Schwarzschild solution. In addition, these results can
be generalized to a kind of more general geometric flow.

\end{abstract}


\maketitle \markright{Fu-Wen Shu}

 {\sl Introduction.} The geometric
flows provide new ways to address a variety of non-linear problems
in Riemannian geometry. In addition, it is shown that they also play
an important role in physics, especially Einstein equation(it is
included as the limiting case of no flow) in general relativity, and
Ricci flow \cite{hamilton} in the renormalization group analysis of
non-linear sigma models in string theory\cite{friedan,leigh} and
various models of boundary quantum field theory of current
interest\cite{lvz}. Recently, Graf \cite{graf} showed that the
Einstein gravity is a limiting case of the Ricci flow gravity. After
that, Headrick and Wiseman \cite{hw} found that the Ricci flow can
be used to study black holes. As to other things, Perelman
\cite{perelman1,perelman2} applied the Ricci flow to prove
Thurston's geometrization conjecture \cite{thurston} concerning the
classification of 3-manifolds (for reviews see \cite{morgan}).

Recently Kong and Liu \cite{kl1} introduced a kind of new flow,
which they called the \textit{hyperbolic geometric flow}. It is a
system of non-linear evolution partial differential equations of
second order. In their remarkable work on hyperbolic flow
\cite{kl1}, they showed that ``\textit{the hyperbolic geometric flow
is a natural and powerful tool to study some problems arising form
differential geometry such as singularities, existence and
regularity}''.

Motivated by these amazing progress of the geometric flows both in
mathematics and physics, we study the gravitational properties of
the geometric evolution equations. Thanks to recent introduction of
the hyperbolic geometric flow, we have a set of general evolution
equations including the Einstein equations (no flow case), the Ricci
flow, and the hyperbolic flow. These equations are \cite{kl1} \be
\alpha_{\mu\nu}\frac{\partial^2 g_{\mu\nu}}{\partial
t^2}+\beta_{\mu\nu}\frac{\partial g_{\mu\nu}}{\partial
t}+\gamma_{\mu\nu}g_{\mu\nu}+2R_{\mu\nu}=0. \label{GE} \ee where
$\alpha_{\mu\nu}$, $\beta_{\mu\nu}$, $\gamma_{\mu\nu}$ are certain
smooth functions (may depend on $t$) on $n$--dimensional complete
Riemannian manifold $\mathcal{M}$, and $g_{\mu\nu}$ is Riemannian
metric, $R_{\mu\nu}$ is Ricci tensor. Obviously, in the case of
$\alpha_{\mu\nu}=\beta_{\mu\nu}=0$ and $\gamma_{\mu\nu}=constant$,
Eq. (\ref{GE}) corresponds to the Einstein equation with regard to
vacuum spacetime; the case of $\alpha_{\mu\nu}=0$,
$\beta_{\mu\nu}=1$ and $\gamma_{\mu\nu}=0$ corresponds to the Ricci
flow, which is analogous to the heat equation in physics; and the
case of $\alpha_{\mu\nu}=1$, $\beta_{\mu\nu}=\gamma_{\mu\nu}=0$
corresponds to the hyperbolic geometric flow, which is analogous to
the wave equation in physics, describing the wave character of the
metrics and curvatures of manifolds.

This letter is an attempt to apply the geometric evolution equation
(\ref{GE}) to study some black-hole problems. Our results show that,
under certain conditions, the Birkhoff theorem\footnote{In fact,
recently Deser\cite{sd} pointed out that this theorem was first
discovered by Jebsen \cite{jebsen}, instead of Birkhoff.}
\cite{birkhoff}
 is still reserved in the frame of the geometric flow
discussed here. We also show that for the spherically symmetric
metric in the $4$--dimensional Riemannian manifold, the Einstein
equation, the Ricci flow, and the hyperbolic geometric flow in
vacuum spacetime all have the same black-hole solutions. In
addition, we can generalize our results to a kind of more general
geometric flow.

\vspace{1mm}

{\sl Birkhoff theorem.} In 1923, Birkhoff \cite{birkhoff,jebsen}
showed that any spherically symmetric solution of the vacuum
Einstein field equations must be static and asymptotically flat. In
this section we will try to extend the Birkhoff theorem to the case
including geometric evolution (\ref{GE}).

\textbf{Theorem 1}. \textit{For the geometric flow (\ref{GE}) in the
$4$--dimensional Riemannian manifold, if $\alpha_{\mu\nu}$,
$\beta_{\mu\nu}$, and $\gamma_{\mu\nu}$ are independent of $t$ and
$r$, then Eq. (\ref{GE}) satisfies the Birkhoff theorem.}

\textbf{Proof}. Consider a line element with spherically symmetric
metric
$$
g_{\mu\nu}=diag \left(B(r,t), -A(r,t), -r^2, -r^2 \sin^2
\theta\right),
$$
 where `` $diag$ '' represents a diagonal matrix. Then we
get,
\begin{eqnarray}
 \begin{cases}
  R_{rr}=\frac{B^{\prime\prime}}{2B}-\frac{B^{\prime}{}^2}{4B^2}-\frac{A^{\prime}B^{\prime}}{4AB}
  -\frac{A^{\prime}}{Ar}-\frac{\ddot{A}}{2B}+\frac{\dot{A}\dot{B}}{4B^2}+\frac{\dot{A}^2}{4AB},\\
  R_{\theta\theta}=-1+\frac1{A}-\frac{rA^{\prime}}{2A^2}+\frac{rB^{\prime}}{2AB},\\
  R_{\varphi\varphi}=R_{\theta\theta} \sin ^2 \theta,\\
  R_{tt}=-\frac{B^{\prime\prime}}{2A}+\frac{A^{\prime}B^{\prime}}{4A^2}
  -\frac{B^{\prime}}{Ar}+\frac{B^{\prime}{}^2}{4AB}+\frac{\ddot{A}}{2A}-\frac{\dot{A}\dot{B}}{4AB}-\frac{\dot{A}^2}{4A^2},\\
  R_{rt}=-\frac{\dot{A}}{Ar},
 \end{cases}\label{riemann}
\end{eqnarray}
where `` $\prime$ '' and `` $\cdot$ '' represent derivatives with
respect to $r$ and $t$, respectively. On account of the fact that
$g_{\mu\nu}$ is diagonal, from (\ref{GE}) we directly have \be
R_{rt}=\dot{A}=0, \label{a}\ee From here we learn that $A(r,t)$ is
independent of $t$, so we let $A(r,t)\equiv a(r)$. Plugging
(\ref{a}) into formula (\ref{riemann}), and letting $B(r,t)\equiv
\rho(t)b(r)$, from (\ref{GE}) one has
\begin{eqnarray}
 \begin{cases}
  \alpha_{00}\ddot{\rho}b+\beta_{00}\dot{\rho}b+\gamma_{00}\rho b=-2\rho\left(-\frac{b^{\prime\prime}}{2a}
  +\frac{a^{\prime}b^{\prime}}{4a^2}-\frac{b^{\prime}}{ar}+\frac{b^{\prime}{}^2}{4ab}\right),\\
  \gamma_{11}a=-2\left(\frac{b^{\prime\prime}}{2b}-\frac{b^{\prime}{}^2}{4b^2}-\frac{a^{\prime}b^{\prime}}{4ab}
  -\frac{a^{\prime}}{ar}\right),\\
  \gamma_{22}r^2=-2\left(-1+\frac1{a}-\frac{ra^{\prime}}{2a^2}+\frac{rb^{\prime}}{2ab}\right).
 \end{cases}\label{decouple}
\end{eqnarray}
The first formula in (\ref{decouple}) can be decoupled as
\begin{eqnarray}
\lambda&=&\alpha_{00}\frac{\ddot{\rho}}{\rho}+\beta_{00}\frac{\dot{\rho}}{\rho}+\gamma_{00} \label{tterm}\\
&=&-2\left(-\frac{b^{\prime\prime}}{2ab}
  +\frac{a^{\prime}b^{\prime}}{4a^2b}-\frac{b^{\prime}}{abr}+\frac{b^{\prime}{}^2}{4ab^2}\right),\label{rterm}
\end{eqnarray}
where $\lambda$ is a constant. From (\ref{tterm}), one can easily
obtain the explicit expression of $\rho(t)$. In the end, we can
rescale the time coordinate as
$$
dt^{\prime}{}^2\equiv \rho(t)dt^2.
$$
In this way, we obtain
$$
ds^2=b(r)dt^{\prime}{}^2+a(r)dr^2+r^2d\Omega^2,
$$
with $a(r)$ and $b(r)$ independent of $t$. This leads to the
Birkhoff theorem in general relativity.

In fact, from the mathematical point of view, we can extend Eq.
(\ref{GE}) to the following more general evolution equations
\begin{eqnarray} \nonumber\alpha^n_{\mu\nu}\frac{\partial^n
g_{\mu\nu}}{\partial t^n}+\alpha^{n-1}_{\mu\nu}\frac{\partial^{n-1}
g_{\mu\nu}}{\partial
t^{n-1}}+\cdots\\
+\alpha^{1}_{\mu\nu}\frac{\partial g_{\mu\nu}}{\partial
t}+\alpha^{0}_{\mu\nu}g_{\mu\nu}+2R_{\mu\nu}=0, \label{general}
\end{eqnarray}
where $\alpha^{n}_{\mu\nu}$ are certain smooth functions (may depend
on $t$) on $m$--dimensional complete Riemannian manifold
$\mathcal{M}$, and $n$ is integer. Note that for the case of $n>2$,
we have little knowledge of its physical counterpart of Eq.
(\ref{general}). Similarly, one can easily obtain a more general
theorem:

\textbf{Theorem 2}. \textit{For the geometric flow (\ref{general})
in the $4$--dimensional Riemannian manifold, if $\alpha^n_{\mu\nu}$
are independent of $t$ and $r$, then Eq. (\ref{general}) satisfies
the Birkhoff theorem.}

One can prove Theorem 2 by using the same technique addressed above
for Theorem 1.

\vspace{1mm}

{\sl Exact black-hole solutions of geometric evolution equation.} As
we have learned in general relativity, if the matter distributes
spherically inside a ball, and the outside of the ball is vacuum,
then we have an exact solution of the Einstein field equation, which
is now called Schwarzschild solution. In this section, we will show
that not only the Einstein equation has black-hole solution, the
Ricci evolution equation and the hyperbolic evolution equation also
have. In order to see this in detail, we let $\gamma_{\mu\nu}=0$ in
Eq. (\ref{GE}), so that it corresponds to Ricci flow as
$\alpha_{\mu\nu}=0$, $\beta_{\mu\nu}=1$, and hyperbolic flow as
$\alpha_{\mu\nu}=1$, $\beta_{\mu\nu}=0$, respectively. In this case,
when taking Eq. (\ref{rterm}) into account, Eq. (\ref{decouple})
becomes
\begin{eqnarray}
 \begin{cases}
  -\frac{b^{\prime\prime}}{2ab}
  +\frac{a^{\prime}b^{\prime}}{4a^2b}-\frac{b^{\prime}}{abr}+\frac{b^{\prime}{}^2}{4ab^2}=-\frac12 \lambda,\\
  \frac{b^{\prime\prime}}{2b}-\frac{b^{\prime}{}^2}{4b^2}-\frac{a^{\prime}b^{\prime}}{4ab}
  -\frac{a^{\prime}}{ar}=0,\\
  -1+\frac1{a}-\frac{ra^{\prime}}{2a^2}+\frac{rb^{\prime}}{2ab}=0.
 \end{cases}
\end{eqnarray}
Solving it, one gets \be y^{\prime}r+y-1+\Lambda r^2=0,
\label{solution}\ee where $y\equiv \frac 1a$ and
$\Lambda=\frac{\lambda}4$.

Integrating Eq. (\ref{solution}), we have
$$
y=1-\frac{2m}r-\frac13\Lambda r^2,$$ where $m$ is an integration
constant (one can easily confirm that $m=\frac{GM}{c^2}>0$ is the
mass of the Schwarzschild black hole). Consequently, we obtain
\begin{eqnarray}
a&=&\left(1-\frac{2m}r-\frac13\Lambda r^2\right)^{-1},\\
\frac{b^{\prime}}{b}&=&2\Lambda ar-\frac{a^{\prime}}a.\label{b}
\end{eqnarray}

\underline{Case 1} $\Lambda=0$: In this case, integrating
Eq.(\ref{b}) directly, one has $b=\frac{\varepsilon}a$, where
$\varepsilon$ is an integration constant which can be set to $1$. In
the end, one has
\begin{equation*}
ds^2=\left(1-\frac{2m}r\right)c^2dt^2-
\left(1-\frac{2m}r\right)^{-1}dr^2+r^2d\Omega^2.
\end{equation*}
This is just the Schwarzschild metric!

\underline{Case 2} $\Lambda\neq 0$: In this case, we rewrite
$\frac {1}{a}$ as
$$
1-\frac{2m}r-\frac{\Lambda}3r^2=-\frac{\Lambda(r-r_a)(r-r_b)(r-r_c)}{3r},
$$
where $r_a$, $r_b$ and $r_c$ are three roots of $\frac{1}{a}=0$,
which depend on the value of the decoupled constant $\Lambda$.
Plugging this formula into Eq.(\ref{b}), and integrating it, we have
$$
b=\varepsilon
(r-r_a)^{\sigma_a}(r-r_b)^{\sigma_b}(r-r_c)^{\sigma_c}\cdot \frac
1a,
$$
where $\varepsilon$ is an integration constant which can be set to
1, and
$$
\sigma_i=\frac{\Lambda r_i}{\kappa_i},
$$
where $\kappa_i\equiv \frac12 \frac{d(1/a)}{dr}\mid_{r=r_i}$, which
corresponds to the surface gravity of the black hole. In the end, we
obtain
\begin{eqnarray*}
ds^2&=&u(r)\left(1-\frac{2m}r-\frac{\Lambda}3r^2\right)c^2dt^2\\&-&
\left(1-\frac{2m}r-\frac{\Lambda}3r^2\right)^{-1}dr^2+r^2d\Omega^2,
\end{eqnarray*}
where
$$
u(r)=(r-r_a)^{\sigma_a}(r-r_b)^{\sigma_b}(r-r_c)^{\sigma_c}.
$$

(a) $\frac1{\Lambda} \geq \frac 94 r_g^2 $, where $r_g=2m$ is the
Schwarzschild radius: In this case, $r_a$, $r_b$, $r_c$ are real.
Let $r_a<r_b<r_c$. After some mathematical analyses, one finds that
$r_a<0$ and $r_c\geq r_b>0$. Consequently, black hole in this case
has two horizons, one of them is the event horizon.

(b) $\frac1{\Lambda} < \frac 94 r_g^2 $: The black hole only has one
horizon--- the event horizon (here we let $r_a$ be the event
horizon). $r_b=r^{*}_c$ are complex in this case.

Surprisingly, as we solve the exact solutions of the more general
geometric flows (\ref{general}), we find that we get the completely
same solutions as done above. This implies that \textit{any
$4$--dimensional spherically symmetric metric in vacuum spacetime
will inevitably form static black holes, as long as it keeps
unchanged or changes isotropically}. This leads to a conjecture,
namely, \textit{any geometric flow which satisfies the Birkhoff
theorem may have a black-hole solution}.

\vspace{1mm}

{\sl Conclusions.} We have shown that for the geometric flows
(\ref{GE}) in the $4$--dimensional Riemannian manifold, including
the Einstein equations(no flow), the Ricci flow, and the hyperbolic
geometric flow, if $\alpha_{\mu\nu}$, $\beta_{\mu\nu}$, and
$\gamma_{\mu\nu}$ are independent on $t$ and $r$, then Eq.
(\ref{GE}) satisfies the Birkhoff theorem, namely, any spherically
symmetric solution of the vacuum geometric evolution equations must
be static and asymptotically flat. Moreover, we can extend the
Birkhoff theorem to the more general flow (\ref{general}), under the
condition that $\alpha^n_{\mu\nu}$ are independent of $r$ and $t$.
This implies that \textit{the spherically symmetric change of the
initial spherically symmetric metric in the vacuum spacetime does
not induce pure gravitational radiation, regardless of the way of
this change}.

As to the exact solution of the geometric evolution equations, our
results have shown that in the $4$--dimensional case of matter
distributed spherically inside a ball, and vacuum outside of the
ball, the Einstein equation, the Ricci flow, and the hyperbolic
geometric flow in vacuum spacetime have the same black-hole
solution. Particularly, in the case of $\Lambda=0$, they all give
the Schwarzschild solution.

\noindent {\sl Acknowledgments.} F. W. Shu wishes to thank Kefeng
Liu for informing me about the reference \cite{kl1}. The work was
supported by the National Natural Science Foundation of China under
Grant No. 10573027 and 10663001, and the Foundation of Shanghai
Natural Science Foundation under Grant No. 05ZR14138.


\begin{thebibliography}{9}


\bibitem{hamilton}
 R. Hamilton, J.\ Differential Geom.\ {\bf17} 255-306 (1982).


\bibitem{friedan}
D.\ H.\ Friedan, Ann.\ Phys.\ {\bf163} 318 (1985).


\bibitem{leigh}
 R.\ Leigh, Mod.\ Phys.\ Lett.\ A {\bf4}
 2767 (1989).


\bibitem{lvz}
 S.\ Lukyanov, E.\ Vitchev and A.\ Zamolodchikov, Nucl.\ Phys.\ B {\bf683} 423 (2004).

\bibitem{graf}
 W.\ Graf, arXiv: gr-qc/0602054.

\bibitem{hw}
 M.\ Headrick and T.\ Wiseman arXiv: hep-th/0606086.

\bibitem{perelman1}
  G.\ Perelman, arXiv: math.\ DG/0211159.


\bibitem{perelman2}
  G.\ Perelman, arXiv: math.\ DG/0303109.

\bibitem{thurston}
  W.\ P.\ Thurston, Bull.\ Amer.\ Math.\ Soc.\ {\bf6} 357 (1982).


\bibitem{morgan}
  J.\ W.\ Morgan, Bull.\ Amer.\ Math.\ Soc.\ (N.S.) {\bf42} 57 (2005).

\bibitem{kl1}
  D.\ X.\ Kong and K.\ Liu, http://www.cms.zju.edu.cn\\/UploadFiles/AttachFiles/200682885946597.pdf.



\bibitem{birkhoff}
G.\ D.\ Birkhoff, \textit{Relativity and modern physics}, Cambridge:
Cambridge university press, (1923).


\bibitem{sd}
S.\ Deser, Gen.\ Relativ.\ Gravit.\ {\bf37} 2251 (2005).



\bibitem{jebsen}
J.\ T.\ Jebsen, Gen.\ Relativ.\ Gravit.\ {\bf37} 2253 (2005).


\end{thebibliography}
\end{document}